\documentclass[preprint,12pt]{elsarticle}

\usepackage{amsmath}
\usepackage{amstext}
\usepackage{amssymb}
\usepackage{lineno,hyperref}
\usepackage{makecell}
\usepackage{color}
\usepackage{soul}
\usepackage[utf8]{inputenc}
\usepackage{fancyhdr}

\hypersetup{colorlinks}

\newcommand{\diag}{\operatorname{diag}}

\definecolor{orange}{RGB}{255,165,0}

\newcommand{\eps}{\epsilon}
\newcommand{\veps}{\varepsilon}
\newcommand{\wt}[1]{\widetilde{#1}}
\newcommand{\wh}[1]{\widehat{#1}}

\newcommand{\orcid}[1]{ORCID: \href{https://orcid.org/#1}{#1}}

\modulolinenumbers[5]

\biboptions{sort&compress}

\begin{document}
\begin{frontmatter}

\title{Neutrino oscillation probabilities through the looking glass}

\makeatletter
\hypersetup{pdftitle={\@title}}
\makeatother

\author[IFIC-UV]{Gabriela Barenboim\fnref{fnGB}}
\ead{gabriela.barenboim@uv.es}
\author[Brook]{Peter B.~Denton\fnref{fnPD}}
\ead{pdenton@bnl.gov}
\author[Fermi]{Stephen J.~Parke\fnref{fnSP}}
\ead{parke@fnal.gov}
\author[IFIC]{Christoph A.~Ternes\fnref{fnCT}}
\ead{chternes@ific.uv.es}

\fntext[fnGB]{\orcid{0000-0002-3249-7467}}
\fntext[fnPD]{\orcid{0000-0002-5209-872X}}
\fntext[fnSP]{\orcid{0000-0003-2028-6782}}
\fntext[fnCT]{\orcid{0000-0002-7190-1581}}

\address[IFIC-UV]{Departament de F{\'i}sica Te{\'o}rica and IFIC, Universitat de Val{\`e}ncia-CSIC, E-46100, Burjassot, Spain}
\address[Brook]{Physics Department, Brookhaven National Laboratory, Upton, New York 11973, USA}
\address[Fermi]{Theoretical Physics Department, Fermi National Accelerator Laboratory, P. O. Box 500, Batavia, IL 60510, USA}
\address[IFIC]{Institut de F\'{i}sica Corpuscular (CSIC-Universitat de Val\`{e}ncia), Parc Cientific de la UV, C/ Catedratico Jos\'e Beltr\'an, 2, E-46980 Paterna (Val\`{e}ncia), Spain}

\begin{abstract}
In this paper we review different expansions for neutrino oscillation probabilities in matter in the context of long-baseline neutrino experiments.
We examine the accuracy and computational efficiency of different exact and approximate expressions.
We find that many of the expressions used in the literature are not precise enough for the next generation of long-baseline experiments, but several of them are while maintaining comparable simplicity.
The results of this paper can be used as guidance to both phenomenologists and experimentalists when implementing the various oscillation expressions into their analysis tools. 
\end{abstract}

\begin{keyword}
Neutrino physics\sep Neutrino oscillations in matter
\end{keyword}

\end{frontmatter}

\fancyhf{}
\renewcommand{\headrulewidth}{0pt}
\setlength{\headheight}{28pt}
\rhead{IFIC/19-07\\FERMILAB-PUB-19-009-T}
\thispagestyle{fancy}

\newpage
\tableofcontents

\section{Introduction}
Over the last two decades neutrino oscillation measurements have become increasingly precise and are now entering the precision era.
Most of the current data coming from experiments using neutrinos from the sun, reactors, the atmosphere and particle accelerators can be described in terms of three-neutrino oscillations, which depend on the six oscillation parameters: two mass splittings $\Delta m_{31}^2$, $\Delta m_{21}^2$, three mixing angles $\theta_{12}$, $\theta_{13}$ and $\theta_{23}$ and a CP-violating phase $\delta$. Many of these parameters are measured rather well as of now~\cite{deSalas:2017kay}. 
However, there are some remaining unknowns, as for example the value of the CP-phase $\delta$, the octant of the atmospheric angle ($\sin^2\theta_{23}<0.5$ or $\sin^2\theta_{23}>0.5$) and the neutrino mass ordering ($\Delta m_{31}^2>0$ or $\Delta m_{31}^2<0$). Note however, that from combining oscillation data with data from cosmological observations a 3.5$\sigma$ preference for normal ordering can be obtained~\cite{Gariazzo:2018pei,deSalas:2018bym}. 
There are also some anomalies, which might suggest the existence of a fourth neutrino, see Refs.~\cite{Dentler:2018sju,Gariazzo:2017fdh} for the current status. 
However, here we will focus only on the case of standard three-neutrino oscillations. 

To obtain functions for the oscillation probabilities one has to solve the corresponding Schr\"odinger equation. 
While this is easily done in the case of vacuum oscillations (i.e.~a free Hamiltonian), it becomes very difficult to do it in the presence of matter due to alterations of the oscillation patterns due to the Wolfenstein matter effect~\cite{Wolfenstein:1977ue,Barger:1980tf,Mikheev:1986gs,Mikheev:1986wj}.
However, if one assumes a sufficiently constant matter profile in the trajectory of the neutrino analytic solutions can be found~\cite{Zaglauer:1988gz}. These expressions are very complicated unfortunately and do not permit for a deeper understanding of the phenomenology of three-neutrino oscillations due to the presence of the $\cos\left(\frac13\cos^{-1}\cdots\right)$ term shown later in Eq.~\ref{eq:ZS_ABCS}. 

Therefore to obtain better insights one may try to find simpler analytic expansions around naturally appearing small parameters. 
Some commonly used small parameters are the matter potential $a/\Delta m_{31}^2$~\cite{Arafune:1997hd}, $\sin\theta_{13}$ or $\sin^2\theta_{13}$~\cite{Cervera:2000kp,Asano:2011nj}, and the ratio of mass splittings $\Delta m_{21}^2/\Delta m_{31}^2$ or $\Delta m_{21}^2/\Delta m_{ee}^2$~\cite{Arafune:1997hd,Freund:2001pn,Akhmedov:2004ny,Friedland:2006pi,Minakata:2015gra,Denton:2016wmg,Denton:2018hal}, where $\Delta m_{ee}^2 \equiv \cos^2\theta_{12}\Delta m_{31}^2 + \sin^2\theta_{12}\Delta m_{32}^2$ \cite{Nunokawa:2005nx,Parke:2016joa}. 
In this paper we analyze, with an eye for both precision and computational speed, different expansions and show how accurate they are and how they have aged as the measurement of the oscillation parameters has evolved over the past twenty years.
It is clear that simplicity is an important trait for an approximate expression.
While simplicity may be somewhat in the eye of the beholder, we use computational speed as a rough proxy for simplicity.

We focus on the DUNE \cite{Acciarri:2016crz} experimental conditions of $L=1300$ km and Earth density of $\rho=3$ g/cm$^3$.
We take the matter density to be constant which is a good approximation since effects due to the variability of the density of the Earth are beyond the sensitivity of DUNE \cite{Kelly:2018kmb}.
Our results will also generally apply to other long-baseline experiments such as NOVA \cite{Ayres:2004js}, T2K/HK \cite{Itow:2001ee,Abe:2014oxa}.
We also discuss the second oscillation maximum which is relevant for T2HKK \cite{Abe:2015zbg} and ESSnuSB \cite{Baussan:2013zcy}.

In the next section we review the various expansions we study. In section \ref{sec:comparison} we compare how well they work comparing them among each other and also in comparison to the exact solution. We also check how fast the expansions can be computed in comparison to the exact analytic solution and to the numerical diagonalization process. Finally, section \ref{sec:conclusions} contains our conclusions.

\section{Expansions under consideration}
\label{sec:list of expressions}
In this section we present the expansions we will compare to the exact expression.
We categorize them into three groups based on their forms.
The first is the ``Madrid-like'' group named for the common city for which one expression was first written down.
Other very similar expressions followed and they will be grouped together accordingly.
The next group is the AKT, MP and DMP group.
This is a series of works that performs two flavor rotations and then perturbation theory.
The final group contains the remaining expressions.
Expressions generally drop terms proportional to various smallness parameters including the ratio of $\Delta m^2$'s, $s_{13}$, or the matter potential.

For historical reasons (i.e.~neutrino factory \cite{Albright:2000xi}) many of these expansions have been performed in the channel $\nu_e\rightarrow\nu_\mu$. However, in the context of long-baseline accelerators the most important channel is $\nu_\mu\rightarrow\nu_e$. Therefore we will present our results in this channel. They are related to each other through the T-relation $P(\nu_\mu\rightarrow \nu_e;\delta)=P(\nu_e\rightarrow \nu_\mu;-\delta)$, just switching the sign of the CP-phase\footnote{Note that sending $\delta\to-\delta$ is equivalent to sending $L\to-L$ under the assumption of CPT invariance.}. If one is interested in antineutrinos, namely $P(\overline{\nu}_\mu\rightarrow \overline{\nu}_e;E)=P(\nu_\mu\rightarrow \nu_e;-E)$, one only has to switch the sign of the neutrino energy.
We will focus on neutrinos, but our results generally apply to antineutrinos as well.
Note that matter effects are not very important in the disappearance channels $\nu_\mu\rightarrow\nu_\mu$ and therefore focusing only on the appearance channel will also not affect the main message of this paper.

While the choice of notation does not affect the precision of these formulas, it does affect their general clarity and overall usefulness.
To this end, we have chosen to use uniform notation throughout this article as much as possible with the various terms defined in Tab.~\ref{tab:defn}.
In doing so we have made several simplifying manipulations, in each case maintaining the exact same mathematical expression.
The relationship between the notation used here and the original notation used is mentioned below whenever applicable.
While these definitions represent fairly commonly used definitions in the literature, some differ by factors of two or other slight changes, so care is required when making comparisons.

\begin{table}
\centering
\caption{The various terms used throughout this paper.
Natural units are assumed throughout unless otherwise mentioned.}
\label{tab:defn}
\begin{tabular}{|l|l|}
\hline
$E$               & Neutrino energy\\
$L$               & Baseline\\\hline
$a$               & $2E\sqrt{2}G_Fn_e$\\
$G_F$             & Fermi's constant\\ 
$n_e$             & Electron \# density\\\hline
$s_{ij}$          & $\sin\theta_{ij}$\\
$c_{ij}$          & $\cos\theta_{ij}$\\
$\Delta m^2_{ij}$ & $m_i^2-m_j^2$\\
$\Delta m^2_{ee}$ & $c_{12}^2\Delta m^2_{31}+s_{12}^2\Delta m^2_{32}$\\[1mm]  \hline
$\Delta_{ij}$     & $\Delta m^2_{ij}L/(4E)$\\
$\Delta_x$        & $xL/(4E)$\\\hline
$\eps$            & $\Delta m^2_{21}/\Delta m^2_{ee}$\\
$\veps$           & $\Delta m^2_{21}/\Delta m^2_{31}$\\\hline
$J_r$             & $s_{23}c_{23}s_{13}c_{13}^2s_{12}c_{12}$\\[1mm]  \hline
$\wh{x}$          & $x$ in matter\\
$\wt{x}$          & Approx. $x$ in matter \\
\hline
\end{tabular}
\end{table}

\subsection{The Madrid-like expressions}
In this subsection we list a few nearly identical expressions and discuss their similarities and differences.

\subsubsection{The Madrid expression (2000)}
This expression  was derived in Ref.~\cite{Cervera:2000kp} (by Cervera, Donini, Gavela, Gomez C\'adenas, Hern\'andez, Mena, and Rigolin, Madrid hereafter) and can be written\footnote{In Ref.~\cite{Cervera:2000kp} $\tilde J\equiv8J_r$ is used instead, the definition of $\Delta_{ij}$ differs by a factor of $L/2$, and terms $A\equiv a/(2E)$ and term $B\equiv |b|/(2E)$ are used. While Ref.~\cite{Cervera:2000kp} defines their $b$-like parameter with absolute value signs, we note that they are not necessary since the probability is even in $b$.}
\begin{align}
 P_{\mu e} ={}& 4s_{23}^2s_{13}^2c_{13}^2\left(\frac{\Delta m^2_{31}}{b}\right)^2\sin^2\Delta_b
 +4c_{23}^2s_{12}^2c_{12}^2\left(\frac{\Delta m^2_{21}}{a}\right)^2\sin^2\Delta_a\nonumber \\
 &+8J_r\frac{\Delta m^2_{21}}{a}\frac{\Delta m^2_{31}}{b}\sin\Delta_a\sin\Delta_b\cos\left(\delta+\Delta_{31}\right)\,,
\label{eq:madrid}
\end{align}
where $b\equiv a-\Delta m^2_{31}$.

The form of this expressions suggests the square of two summed amplitudes.
It is not exactly such a sum due to an extra factor of $c_{13}$ in the interference term which provides the correct CP-violating term in vacuum.
Writing Eq.~\ref{eq:madrid} as the sum of two amplitudes has been examined in various forms in Refs.~\cite{Parke:2005ev,Friedland:2006pi,Parke:2006mr,Nunokawa:2007qh}.

There are two ways to correct this.
The first is to drop one of the $c_{13}$'s in the Jarlskog invariant ($J_r$).
Alternatively, if we add in a factor of $c_{13}$ to one of the amplitudes, we recover the Jarlskog in vacuum correctly while still writing the expression as the sum of two amplitudes.
The natural place to put it is on the $a$ (21) term \cite{Parke:2005ev,Parke:2006mr,Nunokawa:2007qh,Minakata:2013hgk} as this reproduces the vacuum expression exactly.
We note that this provides a negligible change to the precision of the equation as we are correcting an already subleading term (the solar term) by a small amount.
On the other hand we can add the $c_{13}$ term to the $b$ (31) term.
Doing so improves the precision of the Madrid expression for neutrinos by about an order of magnitude at the first oscillation maximum, although this effect is due to a lucky cancellation for the parameters used.
The improvement is more modest elsewhere, and for antineutrinos the precision is a bit worse than the Madrid expression.
As such we do not include such an expression in our subsequent analyses.

In addition, considering the previously identified importance of $\Delta m^2_{ee}$ \cite{Parke:2016joa} and the fact that there is no reason to use $\Delta m^2_{31}$ or $\Delta m^2_{32}$ unless both are treated separately, we have also examined how Eq.~\ref{eq:madrid} performs with $\Delta m^2_{31}\to\Delta m^2_{ee}$.
We find that this change results in somewhat better performance in some cases (modest improvement at $E\gtrsim$ few GeV and considerable improvement at and below the second maximum), in the region of interest for DUNE around a few GeV the performance is essentially the same as the Madrid expression.

\subsubsection{The AJLOS(31) expression (2004)}
In Ref.~\cite{Akhmedov:2004ny} (by Akhmedov, Johansson, Lindner, Ohlsson, and Schwetz, AJLOS hereafter) several expressions are introduced each with different expansion parameters.
We label them by the equation numbers in the original paper\footnote{In Ref.~\cite{Akhmedov:2004ny} $\alpha=\veps$, $A=a/\Delta m^2_{31}$, and $\Delta=\Delta_{31}$ are used. We refer to the equation numbers in the version on the arXiv, not in the journal.}.

The first expression (\#31) drops higher order terms proportional to $\veps$ and $s_{13}$.
\begin{align}
P_{\mu e}={}&4s_{13}^2s_{23}^2\left(\frac{\Delta m^2_{31}}{b}\right)^2\sin^2\Delta_b
+4s_{12}^2c_{12}^2c_{23}^2\left(\frac{\Delta m^2_{21}}{a}\right)^2\sin^2\Delta_a\nonumber\\
&+8\frac{J_r}{c_{13}^2}\frac{\Delta m^2_{21}}{a}\frac{\Delta m^2_{31}}{b}\sin\Delta_a\sin\Delta_b\cos(\Delta_{31}+\delta)\,.
\end{align}
We note that up to a factor of $c_{13}^2$ in each the second and third term this expression is otherwise identical to the Madrid expression in Eq.~\ref{eq:madrid}.

\subsubsection{The FL expression (2006)}
Also the authors of Ref.~\cite{Friedland:2006pi} (by Friedland and Lunardini, FL hereafter) write the probability as the sum of two amplitudes. Using our notation\footnote{In Ref.~\cite{Friedland:2006pi} $\Delta_1=2(\Delta_{32}-\Delta_a)$, $\Delta_2=-2\Delta_a$, $G_1=\Delta_{32}\sin2\theta_{13}e^{i\delta}$, and $G_2=-\Delta_{21}\sin2\theta_{12}$ were used.} they obtain 
\begin{equation}
P_{\mu e}=\left|\Delta_{32}e^{i\delta}s_{13}c_{13}s_{23}\frac{e^{2i(\Delta_{32}-\Delta_a)}-1}{\Delta_{32}-\Delta_a}
-\Delta_{21}s_{12}c_{12}c_{23}\frac{e^{-2i\Delta_a}-1}{\Delta_a}
\right|^2\,.
\label{eq:FL}
\end{equation}
Note that this expression was derived actually in the context of Non-Standard neutrino Interactions (NSI) and that it reduces to eq.~\ref{eq:FL} once all the NSI parameters are set to zero.
This expression is identical to Eq. \ref{eq:madrid} up to using $\Delta m^2_{32}$ instead of $\Delta m^2_{31}$ and factors of $c_{13}$.

\subsection{The AKT, MP and DMP expressions}
In \cite{Agarwalla:2013tza,Minakata:2015gra,Denton:2016wmg} a different technique was used.
Two-flavor rotations were performed to simply diagonalize the Hamiltonian by focusing on the largest off-diagonal terms first.
This means that all channels ($\nu_\alpha\to\nu_\beta$ for $\alpha,\beta\in\{e,\mu,\tau\}$) are handled simultaneously.
In AKT the focus is on the vacuum mass eigenstate basis whereas MP and DMP focus on the flavor basis. This choice affects the order of the two, two-flavor rotations and hence the precision.
Roughly speaking, AKT performs a 12 rotation followed by 13 rotation whereas MP and DMP first perform a 13 rotation followed by a 12 rotation as sketched below.

\subsubsection{The AKT expression (2014)}
\label{sec:AKT}
The authors (Agarwalla, Kao, Takeuchi) of Ref.~\cite{Agarwalla:2013tza} (AKT hereafter) perform the rotations going from largest contribution in the Hamiltonian to smallest in the mass basis.
They begin with the 12 rotation, followed by the 23 rotation and then after commuting with the 12 rotation, they absorb this 23 rotation into a 13 rotation.

Using this approach the effective mixing angles can be written as
\begin{align}
\tan 2\wt{\theta}_{12} & = \frac{\Delta m^2_{21}\sin 2\theta_{12}}{\Delta m^2_{21}\cos 2\theta_{12} - a c_{13}^2 }\,,\nonumber\\ 
\tan 2\wt{\theta}_{13} & = \frac{\Delta m^2_{ee}\sin 2\theta_{13}}{\Delta m^2_{ee}\cos 2\theta_{13}-a}\,.
\end{align}
Note that $\delta$ and $\theta_{23}$ are treated as constant in matter. The eigenvalues of the Hamiltonian are now given by
\begin{align}
\lambda_{1/p} & = \frac{ (\Delta m^2_{21}+a c_{13}^2)\, (-/+) \, \sqrt{ (\delta m^2_{21}-a c_{13}^2)^2 + 4 a c_{13}^2 s_{12}^2 \Delta m^2_{21} }}{ 2 }\,,\\
\lambda_{2/3} & = \frac{ \left( \lambda_{p} + \Delta m^2_{31}+a s_{13}^2 \right) \,(-/+)\, \sqrt{ \left( \lambda_{p} - \Delta m^2_{31}-a s_{13}^2 \right)^2 + 4 a^2 s^2_{\widetilde{12}}\,c_{13}^2\, s_{13}^2 } }{ 2 } \,.\nonumber
\end{align}
From this expressions we obtain the mass splittings in matter simply via the relation $\Delta \wt{m^2}_{ij} = \lambda_{i}-\lambda_{j}$. The oscillation probability can now be obtained by replacing the vacuum parameters with the matter parameters in the vacuum oscillation probability, see Eqs.~\ref{eq:vacuum probability} and \ref{eq:vacuum coefficients} below.
While $\Delta m^2_{ee}$ was not explicitly used in Ref.~\cite{Agarwalla:2013tza}, it appears in several places nonetheless and we made the substitution here for simplicity.

\subsubsection{The MP expression (2015)}
\label{sec:MP}
After one rotation in the 13 sector, we have an expression that is an expansion in $\eps c_{12}s_{12}$.
This is the expression in Ref.~\cite{Minakata:2015gra} (by Minakata and Parke, MP hereafter).
The expression is
\begin{align}
P_{\mu e} = {}&\left\{s^2_{23} \sin^2 2 \theta_{13}+ 4 \epsilon J_r \cos \delta 
\left[ \frac{ (\lambda_{+} - \lambda_{-}) - ( \Delta m_{ee}^2 - a ) }{  ( \lambda_{+} - \lambda_{0} ) } \right]\right\}\nonumber\\
&\times \left(\frac{\Delta m_{ee}^2}{ \lambda_{+} - \lambda_{-} }\right)^2 \sin^2 \frac{ (\lambda_{+} - \lambda_{-}) L}{ 4E } \nonumber \\
&+ 8 \eps J_r \frac{ (\Delta m_{ee}^2)^3 }{ ( \lambda_{+} - \lambda_{-} ) ( \lambda_{+} - \lambda_{0} ) ( \lambda_{-} - \lambda_{0} ) }\nonumber\\
&\times
\sin \frac{ (\lambda_{+} - \lambda_{-}) L}{ 4E } 
\sin \frac{ (\lambda_{-} - \lambda_{0}) L}{ 4E}
\cos \left( \delta + \frac{ (\lambda_{+} - \lambda_{0}) L}{ 4E } \right)\,.
\label{eq:P-emu-sec3}
\end{align}
where\footnote{In Ref.~\cite{Minakata:2015gra} $\Delta m^2_{\rm ren}=\Delta m^2_{ee}$ is used and the definition of the eigenvalues is shifted by $\eps\Delta m^2_{ee}s_{12}^2$.}
\begin{align} 
\lambda_{0} &=\eps   \cos 2\theta_{12}  \Delta m_{ee}^2\,,\quad 
\lambda_{\pm} = 
\frac{ 1 }{ 2 } \left[ \Delta m_{ee}^2 + a  \pm \Delta \wt{m^2}_{ee} \right]\,,
\label{eq:lambda-pm0}\\
{\rm with} & \quad  \Delta \wt{m^2}_{ee} \equiv \Delta m_{ee}^2\sqrt{(\cos2\theta_{13}-a/\Delta m_{ee}^2)^2+\sin^22\theta_{13}}=\lambda_+-\lambda_-\,.\nonumber
\end{align}
This expression is very accurate except near the solar resonance when ${\Delta_{21}>1}$.

\subsubsection{The DMP expression (2016) }
To address the solar resonance, an additional rotation was performed in Ref.~\cite{Denton:2016wmg} (by Denton, Minakata, and Parke, DMP hereafter).
The order of the rotations (12 then 13 after an initial constant 23 rotation) is chosen to diagonalize the largest remaining off-diagonal term at each step.
This procedure also removes both level crossings with the minimal number of new angles.
After these two rotations, perturbation theory is now possible everywhere, although the zeroth order expression (DMP$^0$) is sufficiently precise for future long-baseline experiments.

Here, as in AKT in section \ref{sec:AKT} above, the authors do not derive formulas for the oscillation probabilities directly, but rather for the oscillation parameters in matter and then write the probability 
\begin{equation}
\wt P_{\alpha\beta}(\Delta m^2_{ij},{\theta}_{ij},{\delta})=P_{\alpha\beta}(\Delta \wt{m^2}_{ij},\wt\theta_{ij},\wt\delta)\,,
\end{equation}
where $\wt x$ refers to the approximate expression for the quantity $x$ evaluated in matter.
That is, to an excellent approximation the oscillation probability in matter has the same form as the expression in vacuum with only $\Delta m^2_{21}$, $\Delta m^2_{31}$, $\theta_{12}$, and $\theta_{13}$ replaced by their approximate matter equivalents ($\wt\theta_{23}$ and $\wt\delta$ are roughly constant in matter).
The authors obtain to zeroth order the following expressions\footnote{In Ref.~\cite{Denton:2016wmg} $\phi=\wt\theta_{13}$, ${\psi}=\wt\theta_{12}$, $\lambda_i=\wt{m^2}_i$ were used. The current notation is consistent with Ref.~\cite{Denton:2018cpu}. The following expression from \cite{Denton:2016wmg} is also useful: $\cos^2(\theta_{13}-\wt\theta_{13})=c_{\wt{13}}^2c_{13}^2+s_{\wt{13}}^2s_{13}^2+\sin2\wt\theta_{13}c_{13}s_{13} = (\Delta \wt{m^2}_{ee}+\Delta m^2_{ee}-a \cos 2\theta_{13})/(2\Delta \wt{m^2}_{ee})$.},
\begin{align}
\sin^2 \wt{\theta}_{13} &=\frac{1}{2}\left(1-\frac{\Delta m_{ee}^2\cos2\theta_{13}-a}{\Delta \wt{m^2}_{ee}} \right)\,,    \nonumber\\ 
\Delta \wt{m^2}_{ee}&=\Delta m_{ee}^2\sqrt{(\cos2\theta_{13}-a/\Delta m_{ee}^2)^2+\sin^22\theta_{13}}\,,\nonumber\\
\sin^2\wt{\theta}_{12} &=\frac{1}{2}\left(1-\frac{\Delta m_{21}^2\cos2\theta_{12}-a_{12}}{\Delta \wt{m^2}_{21}} \right)\,,  \nonumber\\
&  {\rm where}  \quad a_{12} = \frac{1}{2} \, (a+\Delta m_{ee}^2-\Delta \wt{m^2}_{ee})  \,,\nonumber\\
\Delta \wt{m^2}_{21}&=\Delta m_{21}^2\sqrt{(\cos2\theta_{12}-a_{12}/\Delta m_{21}^2)^2+\cos^2(\theta_{13}-\wt{\theta}_{13})\sin^22\theta_{12}}\,,\nonumber\\
\Delta \wt{m^2}_{31}&=\Delta m_{31}^2+\frac{a}{4}+\frac{1}{2} \left(\Delta \wt{m^2}_{21}-\Delta m_{21}^2 \right)+\frac{3}{4}\left(\Delta \wt{m^2}_{ee} - \Delta m_{ee}^2\right)\,.\label{eq:DMP0}
\end{align}
Note that we have $\tilde{\delta}=\delta$ and $\tilde{\theta}_{23}=\theta_{23}$ at zeroth order.
The expansion parameter is
\begin{equation}
\eps'\equiv\eps\sin(\wt\theta_{13}-\theta_{13})s_{12}c_{12}<0.015\,,
\label{eq:epsp}
\end{equation}
and is zero in vacuum confirming that this expression returns the exact expression in vacuum.

We recall the vacuum expression here, which we write in the following form,
\begin{equation}
P_{\mu e}=4\mathbb C_{21}\sin^2\Delta_{21}+4\mathbb C_{31}\sin^2\Delta_{31}+4\mathbb C_{32}\sin^2\Delta_{32}+8\mathbb D\sin\Delta_{21}\sin\Delta_{31}\sin\Delta_{32}\,,
\label{eq:vacuum probability}
\end{equation}
where the coefficients are,
\begin{align}
\mathbb C_{21}&=c_{13}^2s_{12}^2c_{12}^2(c_{23}^2-s_{13}^2s_{23}^2)+\cos2\theta_{12}J_r\cos\delta\,,\nonumber\\
\mathbb C_{31}&=s_{13}^2c_{13}^2c_{12}^2s_{23}^2+J_r\cos\delta\,,\nonumber\\
\mathbb C_{32}&=s_{13}^2c_{13}^2s_{12}^2s_{23}^2-J_r\cos\delta\,,\nonumber\\
\mathbb D&=-J_r\sin\delta\,.
\label{eq:vacuum coefficients}
\end{align}
DMP$^0$ is then Eqs.~\ref{eq:vacuum probability}, \ref{eq:vacuum coefficients} where the vacuum parameters are replaced with the approximate matter ones given in Eq.~\ref{eq:DMP0}.

Also note that $\Delta\wt{m^2}_{ee}$ (which was further explored in \cite{Denton:2018cpu}) is the same as $\lambda_+-\lambda_-$ in Eq.~\ref{eq:lambda-pm0} above from the MP formula in \ref{sec:MP}.
Successive orders of precision can also be calculated by following perturbation theory in a straightforward fashion as is done through second order in \cite{Denton:2016wmg} with compact expressions provided through first order or by correcting the mixing angles directly \cite{Denton:2018fex}.
Here we focus on zeroth and first orders only (DMP$^0$ and DMP$^{1}$ hereafter respectively) as they are already extremely precise.
Successive orders add $\sim2.5$ additional orders of magnitude of precision if desired and expressions through second order exist in \cite{Denton:2016wmg}.

The first order corrections to the coefficients from Eq.~\ref{eq:vacuum coefficients} are given by the following expressions,
\begin{align}
\mathbb C_{21}^{(1)}&=\eps'\Delta m^2_{ee}\left(\frac{F_1}{\Delta\wt{m^2}_{31}}+\frac{F_2}{\Delta\wt{m^2}_{32}}\right)\,,\nonumber\\
\mathbb C_{31}^{(1)}&=\eps'\Delta m^2_{ee}\left(\frac{F_1+G_1}{\Delta\wt{m^2}_{31}}-\frac{F_2}{\Delta\wt{m^2}_{32}}\right)\,,\nonumber\\
\mathbb C_{32}^{(1)}&=\eps'\Delta m^2_{ee}\left(-\frac{F_1}{\Delta\wt{m^2}_{31}}+\frac{F_2+G_2}{\Delta\wt{m^2}_{32}}\right)\,,\nonumber\\
\mathbb D^{(1)}&=\eps'\Delta m^2_{ee}\left(\frac{K_1}{\Delta\wt{m^2}_{31}}-\frac{K_2}{\Delta\wt{m^2}_{32}}\right)\,,
\end{align}
where $\eps'\equiv\eps\sin(\wt\theta_{13}-\theta_{13})s_{12}c_{12}$ as shown in Eq.~\ref{eq:epsp} above, and
\begin{align}
F_1&=c_{\wt{13}} s^2_{\wt{12}}[s_{\wt{13}} s_{\wt{12}} c_{\wt{12}}(c^2_{23}+\cos2\wt\theta_{13}s^2_{23})-s_{23}c_{23}(s^2_{\wt{13}} s^2_{\wt{12}}+\cos2\wt\theta_{13}c^2_{\wt{12}})\cos\delta]\,,\nonumber\\
G_1&=-2s_{\wt{13}} c_{\wt{13}} s_{\wt{12}}(s^2_{23}\cos2\wt\theta_{13}c_{\wt{12}}-s_{23}c_{23}s_{\wt{13}} s_{\wt{12}}\cos\delta)\,,\nonumber\\
K_1&=-s_{23}c_{23}c_{\wt{13}} s^2_{\wt{12}}(c^2_{\wt{13}} c^2_{\wt{12}}-s^2_{\wt{13}})\sin\delta\,,
\end{align}
and the $F_2,G_2,K_2$ expressions are related to the above by making the transformation $c_{\wt{12}}^2\leftrightarrow s_{\wt{12}}^2$, $c_{\wt{12}}s_{\wt{12}}\to-c_{\wt{12}}s_{\wt{12}}$, and $m_1\leftrightarrow m_2$.
This correction is DMP$^1$.
Note that the expressions of Eq.~\ref{eq:vacuum coefficients} are also invariant under this transformation \cite{Denton:2016wmg}.

\subsection{Other expressions}
Here we list other expressions in the literature that do not fall into the above two categories.

\subsubsection{The AKS expression (1999)}
The oldest expression we consider in this paper is the one derived in Ref.~\cite{Arafune:1997hd} (by Arafune, Koike, and Sato, AKS hereafter). Here the authors obtain 
\begin{align}
P_{\mu e}={}&
4 \sin^2 \Delta_{31} c_{13}^2 s_{13}^2 s_{23}^2
\left(  1 + \frac{2 a}{\Delta m^2_{31}}\cos2\theta_{13} \right)
\nonumber \\
&+4 \Delta_{31} \sin (2\Delta_{31}) c_{13}^2 s_{13} s_{23} \nonumber \\
&\times\left\{ - \frac{a}{\Delta m^2_{31}} s_{13} s_{23} \cos2\theta_{13}  + \veps s_{12}    (-s_{13} s_{23} s_{12} + \cos\delta c_{23} c_{12}) \right\} \nonumber \\
&- 8J_r \Delta_{21} \sin^2 \Delta_{31} \sin\delta\,.
\end{align}

\subsubsection{The MF expression (2001)}
\label{sec:MF}
In Ref.~\cite{Freund:2001pn} (by Freund, MF hereafter) the author separates the oscillation probability\footnote{We use Eq.~36 of Ref.~\cite{Freund:2001pn} since Eq.~38 behaves poorly for large values of $\theta_{13}$.} in sub-terms, $P_{\mu e} = P_0 + P_{\sin\delta}+ P_{\cos\delta}+ P_1 + P_2 + P_3$. These terms are given by\footnote{In Ref.~\cite{Freund:2001pn} $\hat{C}=C_{13}$ is used and there is a typo wherein $\cos\theta_{13}^2$ is written instead of $\cos^2\theta_{13}$.}
\begin{align}
P_0 ={}&  \frac{4s^2_{23} s_{13}^2c_{13}^2}{C_{13}^2} \sin^2(\Delta_{31} C_{13} )\, , \nonumber \\
P_{\sin\delta} ={}&  -4\sin\delta \frac{\Delta m^2_{21}}{a} \frac{ s_{12}c_{12} s_{13} s_{23}c_{23}}{C_{13}} \sin( C_{13}  \Delta_{31})\nonumber \\
 &\times  \left[ \cos( C_{13} \Delta_{31}) - \cos(\Delta_{31}+\Delta_a) \right] \, , \nonumber \\
P_{\cos\delta} ={}&  -4 \cos\delta \frac{\Delta m^2_{21}}{a} \frac{ s_{12}c_{12} s_{13} s_{23}c_{23}}{C_{13}} \sin( C_{13}  \Delta_{31})\, , \nonumber \\
 &\times \left[ \sin( C_{13} \Delta_{31}) - \sin(\Delta_{31}+\Delta_a) \right]\, , \nonumber \\
P_1 ={}&  - 4\veps \frac{1-\frac{a}{\Delta m^2_{31}} \cos 2\theta_{13}}{ C_{13}^3} s^2_{12} s^2_{13} c^2_{13} s^2_{23}  \Delta_{31} \sin(2 \Delta_{31}  C_{13} ) \nonumber \\
     & - 4\veps \frac{2\frac{a}{\Delta m^2_{31}}(\frac{a}{\Delta m^2_{31}} - \cos 2\theta_{13})}{ C_{13} ^4} s^2_{12} s^2_{13} c^2_{13} s^2_{23} \sin^2(\Delta_{31}  C_{13} )\, , \nonumber \\
P_2 ={}&  4\frac{\Delta m^2_{21}}{a} \frac{C_{13}  + \frac{a}{\Delta m^2_{31}}\cos 2\theta_{13} - 1}{C_{13} ^2} s_{12} c_{12} s_{13} s_{23} c_{23} \sin^2(\Delta_{31}  C_{13}  )\, , \nonumber \\
P_3 ={}&  8\left(\frac{\Delta m^2_{21}}{a}\right)^2 \frac{C_{13}  c^2_{23} s^2_{12}c^2_{12}}{\cos^2\theta_{13}(C_{13}  + \cos 2\theta_{13} - a/\Delta m^2_{31})} \sin^2\left[\frac{1}{2}(1- C_{13} )\Delta_{31}+\frac{1}{2}\Delta_a\right] \, .
\label{eq:mf}
\end{align}
The expression in Eq.~\ref{eq:mf} is similar to the AJLOS(48) expression in Eq.~\ref{eq:ajlos48} below.

\subsubsection{The AJLOS(48) expression (2004)}
The second AJLOS (Ref.~\cite{Akhmedov:2004ny}) expression (\#48) focuses only on $\veps$ as a smallness parameter.
The first two orders, $P_{\mu e}=P_{\mu e}^{(0)}+\veps P_{\mu e}^{(1)}$, are
\begin{align}
P_{\mu e}^{(0)}={}&4s_{23}^2s_{13}^2c_{13}^2\frac{\sin^2(C_{13}\Delta_{31})}{C_{13}^2}\nonumber\\
P_{\mu e}^{(1)}={}&-8s_{12}^2s_{23}^2s_{13}^2c_{13}^2\frac{\sin(C_{13}\Delta_{31})}{C_{13}^2}\left[\Delta_{31}\frac{\cos(C_{13}\Delta_{31})}{C_{13}}\left(1-\frac{a}{\Delta m^2_{31}}\cos2\theta_{13}\right)\right.\nonumber\\
&\left.-\frac{a}{\Delta m^2_{31}}\frac{\sin(C_{13}\Delta_{31})}{C_{13}}\frac{\cos2\theta_{13}-a/\Delta m^2_{31}}{C_{13}}\right]+4s_{13}s_{12}c_{12}s_{23}c_{23}\nonumber\\
&\times\frac{\Delta m^2_{31}\sin(C_{13}\Delta_{31})}{aC_{13}^2}\left\{\vphantom{\frac{a}{\Delta m^2_{31}}}\sin\delta\left[\cos(\Delta_{31}+\Delta_a)-\cos(C_{13}\Delta_{31})\right]C_{13}\right.\nonumber\\
&\left.+\cos\delta\left[C_{13}\sin(\Delta_{31}+\Delta_a)-\left(1-\frac{a}{\Delta m^2_{31}}\cos2\theta_{13}\right)\sin(C_{13}\Delta_{31})\right]\right\}\,,
\label{eq:ajlos48}
\end{align}
where $C_{13}\equiv\sqrt{\sin^22\theta_{13}+(a/\Delta m^2_{31}-\cos2\theta_{13})^2}$ (this factor is also used in MF in the previous subsection, \ref{sec:MF}). We note, $C_{13}$ is equivalent to $\Delta\wt{m^2}_{ee}/\Delta m^2_{ee}$ from DMP \cite{Denton:2016wmg,Denton:2018cpu} after the change $\Delta m^2_{31}\to\Delta m^2_{ee}$.

There is also a third expression in the AJLOS paper, Eq.~66, but this is designed for the solar sector and does quite poorly for long-baseline oscillations which are dominated by the atmospheric term.
For this reason we do not consider it here.

\subsubsection{The AM expression (2011)}
We study the expression obtained by the authors of Ref.~\cite{Asano:2011nj} (by Asano and Minakata, AM hereafter). Here the authors use as expansion parameters $s_{13}\simeq\sqrt{\Delta m_{21}^2/\Delta m_{31}^2}$. The authors divide their expressions in powers of $s_{13}$, $P_{\mu e}^{(0)}+P_{\mu e}^{(1)}+P_{\mu e}^{(3/2)}+P_{\mu e}^{(2)}+P_{\mu e}^{(5/2)}+\ldots$ Each superscript refers to the order of $s_{13}^2$ and $\Delta m^2_{21}/\Delta m^2_{31}$.
For our channel of interest they obtain\footnote{In Ref.~\cite{Asano:2011nj} $r_\Delta=\veps$ and $\Delta=2\Delta_{31}/L$ are used.}
\begin{align}
P_{\mu e}^{(0)} ={}&0\,,\nonumber\\
P_{\mu e}^{(1)} ={}&4 s^2_{23} s^2_{13} \frac{\sin^2[(1 - r_{A}) \Delta_{31}]}{ (1 - r_{A})^2 }\,, 
\nonumber\\
P_{\mu e}^{(3/2)} ={}&8 J_{r} \frac{ \veps }{ r_{A} (1 - r_{A}) }
\cos \left( \delta + \Delta_{31} \right)
\sin ( r_{A} \Delta_{31} )
\sin [(1 - r_{A}) \Delta_{31}]\,,
\nonumber\\
P_{\mu e}^{(2)} ={}&4 c^2_{23} c^2_{12} s^2_{12} 
\left( \frac{ \veps }{ r_{A} } \right)^2 
\sin^2 ( r_{A} \Delta_{31})\nonumber \\
&- 4 s^2_{23} \left[ s^4_{13} \frac{ (1 + r_{A})^2 }{ (1 - r_{A})^4 } - 2 s^2_{12} s^2_{13} \frac{ \veps r_{A} }{ (1 - r_{A})^3 } \right] 
\sin^2 [(1 - r_{A}) \Delta_{31}]\nonumber \\
&+4 s^2_{23} \left[ 2 s^4_{13} \frac{ r_{A} }{ (1 - r_{A})^3 } - s^2_{12} s^2_{13} \frac{ \veps }{ (1 - r_{A})^2 } \right] 
\Delta_{31} \sin [2(1 - r_{A}) \Delta_{31}]\,,
\label{Pemu-2}
\end{align}
where $r_A\equiv a/\Delta m_{31}^2$. The $5/2$ order term is given by
\begin{align}
P_{\mu e}^{(5/2)} ={}&8 J_{r} s^2_{13} \frac{ \veps r_{A} }{ (1 - r_{A})^3 } 
\cos \delta \sin^2 [(1 - r_{A}) \Delta_{31}]\nonumber \\
&+8 J_{r} \frac{ \veps }{ r_{A} (1 - r_{A}) } 
\left[ - 2 s^2_{13}  \frac{ r_{A} }{ (1 - r_{A})^2 } + (c^2_{12} - s^2_{12}) \frac{ \veps }{ r_{A} } + s^2_{12} \frac{ \veps r_{A} }{ 1 - r_{A} } \right] \nonumber \\
&\times \cos \left( \delta + \Delta_{31} \right)\sin ( r_{A} \Delta_{31})\sin [ (1 - r_{A}) \Delta_{31}]\nonumber \\
&+16 J_{r} s^2_{13}  \frac{ \veps \Delta_{31}}{ (1 - r_{A})^2 }  \cos \left( \delta + \Delta_{31} \right)\sin ( r_{A} \Delta_{31})\cos [ (1 - r_{A}) \Delta_{31}]\nonumber \\
&- 8 J_{r} s^2_{12} \frac{ \veps^2 \Delta_{31}}{ r_{A} (1 - r_{A}) } \cos \left( \delta + r_{A} \Delta_{31} \right) \sin ( r_{A} \Delta_{31})\nonumber \\
&-8 J_{r} c^2_{12} \frac{ \veps^2 \Delta_{31}}{ r_{A} (1 - r_{A}) } \cos \left[ \delta + (1 + r_{A}) \Delta_{31} \right]\sin [ (1 - r_{A}) \Delta_{31}]\nonumber \\
&-8 J_{r} \frac{ \veps \Delta_{31}}{ r_{A} (1 - r_{A}) } \left( s^2_{13} \frac{ r_{A}  }{ 1 - r_{A} } -  s^2_{12} \veps \right)\nonumber\\
&\times\cos \left[ \delta +  (1 - r_{A}) \Delta_{31} \right]\sin [ (1 - r_{A}) \Delta_{31}]\,.
\label{Pemu-5/2}
\end{align}
We will consider the precision and speed both of the expression through second order (AM$^2$ hereafter) and through $5/2$ order (AM$^{5/2}$ hereafter).

\subsection{Exact expressions}
For completeness, we discuss two different means of exactly calculating the oscillation probabilities.
The first uses the analytic solution to a cubic equation and the second involves numerically diagonalizing the Hamiltonian.
We have verified that these are equivalent up to numerical precision, $\sim10^{-13}$.

Several pieces of software designed to solve neutrino oscillations in matter also exist in the literature \cite{Barger:1980tf,Kopp:2006wp,nusquids}.
While these may offer modest improvements in speed over off-the-shelf linear algebra packages, we have verified that they still do not compete with analytic expressions in terms of speed.
In fact, several of them are the same, in part or in full, as the ZS expression in the next section.
NuSquids \cite{nusquids} is a bit different than the other options in that once it has solved the differential equation for a given matter profile and baseline, extracting the probability (or the flux more generally) for a given energy is fairly efficient.
It starts to become computational efficient for long-baseline at $\gtrsim1000$ energy computations \cite{weaver}.

\subsubsection{The ZS expression (1988)}
It is possible to solve the characteristic polynomial of the Hamiltonian in matter directly. Using this approach one can express the eigenvalues in matter in terms of the vacuum parameters which involves solving a completely general cubic equation\footnote{The original solution of the cubic equation was from \cite{cardano} based on work by Scipione del Ferro and Niccolò Fontana Tartaglia in the sixteenth century.}. This was done in Ref.~\cite{Zaglauer:1988gz} (by Zaglauer and Schwarzer, ZS hereafter), where the authors obtain
\begin{align}
\wh{m^2}_1 &= \frac{A}{3}-\frac{\sqrt{A^2-3B}S}{3}-\frac{\sqrt{3}\sqrt{A^2-3B}\sqrt{1-S^2}}{3}\,,\nonumber\\ 
\wh{m^2}_2 &= \frac{A}{3}-\frac{\sqrt{A^2-3B}S}{3}+\frac{\sqrt{3}\sqrt{A^2-3B}\sqrt{1-S^2}}{3}\,,\nonumber\\
\wh{m^2}_3 &= \frac{A}{3}+\frac{2\sqrt{A^2-3B}S}{3}\,.
\end{align}
The mass splittings in matter are then given by 
\begin{align}
\Delta \wh{m^2}_{21} &= \frac{2\sqrt{3}}{3}\sqrt{A^2-3B}\sqrt{1-S^2}\,,\nonumber\\
\Delta \wh{m^2}_{31} &= \sqrt{A^2-3B}S + \frac{\sqrt{3}}{3}\sqrt{A^2-3B}\sqrt{1-S^2}\,,
\end{align}
where $\wh{x}$ refers to the given quantity in matter and\footnote{Note, that in Ref.~\cite{Zaglauer:1988gz} $D=a$ and there are two typos: the root in the denominator of $S$ should be to the $3/2$ power not $1/3$ and the numerator of $e^{-i\wh{\delta}}$ should have a factor of $s_{23}c_{23}$ instead of $s_{23}c_{23}^2$.}
\begin{align}
A &= \Delta m_{21}^2 + \Delta m_{31}^2 +a \,,\nonumber\\
B &= \Delta m_{21}^2\Delta m_{31}^2 + a [\Delta m_{31}^2c_{13}^2+\Delta m_{21}^2(c_{13}^2c_{12}^2+s_{13}^2)]\,,\nonumber\\
C &= a\Delta m_{21}^2\Delta m_{31}^2c_{13}^2c_{12}^2\,,\nonumber\\
S &= \cos\left\{\frac{1}{3}\arccos\left[\frac{2A^3-9AB+27C}{2(A^2-3B)^{3/2}}\right]\right\}\,.
\label{eq:ZS_ABCS}
\end{align}
The parameter $S$ is the one mentioned in the introduction that has $\cos\left(\frac13\cos^{-1}\cdots\right)$ which ends up in all the final parameters: mass squared differences, mixing angles, and the CP-phase.
The mixing angles and the CP-phase are
\begin{align}
s_{\wh{12}}^2 &= \frac{-[(\wh{m^2}_2)^2-\alpha \wh{m^2}_2 +\beta]\Delta \wh{m^2}_{31}}{\Delta \wh{m^2}_{32}[(\wh{m^2}_1)^2-\alpha \wh{m^2}_1 +\beta]-\Delta \wh{m^2}_{31}[(\wh{m^2}_2)^2-\alpha \wh{m^2}_2 +\beta]}\,,\nonumber\\
s_{\wh{13}}^2 &= \frac{(\wh{m^2}_3)^2-\alpha \wh{m^2}_3 +\beta}{\Delta \wh{m^2}_{31}\Delta \wh{m^2}_{32}}\,,\nonumber\\
s_{\wh{23}}^2 &= \frac{E^2s_{23}^2+F^2c_{23}^2+2EFc_{23}s_{23}\cos\delta}{E^2+F^2}\,,\nonumber\\
e^{-i\wh \delta} &= \frac{(E^2e^{-i\delta}-F^2e^{i\delta})c_{23}s_{23}+EF(c_{23}^2-s_{23}^2)}{\sqrt{(E^2s_{23}^2+F^2c_{23}^2+2EFc_{23}s_{23}\cos\delta)(E^2c_{23}^2+F^2s_{23}^2-2EFc_{23}s_{23}\cos\delta)}}\,,
\end{align}
where
\begin{align}
\alpha &= \Delta m_{31}^2c_{13}^2 + \Delta m _{21}^2 (c_{13}^2c_{12}^2+s_{13}^2)\,,\nonumber\\
\beta &= \Delta m_{21}^2\Delta m_{31}^2c_{13}^2c_{12}^2\,,\nonumber\\
E &= [\Delta m_{31}^2(\wh{m^2}_3-\Delta m_{21}^2)-\Delta m_{21}^2(\wh{m^2}_3-\Delta m_{31}^2)s_{12}^2]c_{13}s_{13}\,,\nonumber\\
F &=  \Delta m_{21}^2(\wh{m^2}_3-\Delta m_{31}^2)c_{12}s_{12}c_{13}\,. 
\end{align}
As in the case of DMP the oscillation probabilities in matter can now be obtained by simply replacing the vacuum parameters with the matter parameters in the vacuum oscillation probability from Eq.~\ref{eq:vacuum probability}.

\subsubsection{Numerical diagonalization}
It is also possible to diagonalize the Hamiltonian numerically. The Schr\"odinger equation in matter can be written in matrix form as 
\begin{equation}
 i\frac{d}{dt}\,\Psi_{\alpha}(t)=\mathcal{H}\,\Psi_\alpha(0),
\end{equation}
where 
\begin{equation}
 \Psi_\alpha(t)=
 \begin{pmatrix}
  \psi_{\alpha e}(t) \\ \psi_{\alpha \mu}(t) \\ \psi_{\alpha \tau}(t)
 \end{pmatrix},\quad
\mathcal{H}=\frac{1}{2E}(U\mathbb{M}^2U^\dagger + \mathbb{A}).
\end{equation}
Here $\psi_{\alpha \beta}(t) = \langle \nu_\beta|\nu_\alpha(t)\rangle$ is the oscillation amplitude, $U=U_{23}(\theta_{23}) \times U_{13}(\theta_{13},\delta)$ $\times U_{12}(\theta_{12})$ is the PMNS-matrix and $\mathbb{M}^2=\diag(0,\Delta m_{21}^2,\Delta m_{31}^2)$. The matter potential is given by $\mathbb{A}=\diag(a,0,0)$, where as always in this work $a$ is constant. In principle this equation can be solved easily by writing $\mathcal{H} = R D R^\dagger$, where $D=\diag(d_1,d_2,d_3)$ is the diagonal matrix containing the eigenvalues $d_i$ of $\mathcal{H}$, and $R$ is the diagonalization matrix of eigenvectors. Then the solution is easily found to be 

\begin{equation}
 \Psi_{\alpha}(t) = R\,\diag(e^{-iLd_1},e^{-iLd_2},e^{-iLd_3})\,R^\dagger\, \Psi_{\alpha}(0).
\end{equation}
This diaginalization can be performed numerically using for example the packages \texttt{Eigen}~\cite{Eigen} or \texttt{HEigensystem}~\cite{Hahn:2006hr,Heigensystem}, among others. We will refer to this method as Diag from here on. 

\section{Comparison of expansions}
\label{sec:comparison}
We now compare the usefulness of each expression.
First we describe the behavior of the various expressions under several useful limits.
Next, we define two metrics: precision and speed/simplicity.
Precision can be quantified as either the error or fractional error between a given expression and the exact expression.
We focus on fractional error since that is more relevant for experiments although it can become misleading when the probability is small or goes to zero such as in the high energy limit.
While the simplicity of an expression is a somewhat subjective metric, the computational efficiency is somewhat more scientific and quantitative.

\subsection{Expansion term}
Each approximate expression is an expansion in one or more parameters.
In order to clearly show how each expression behaves, Tab.~\ref{tab:expansion} shows which parameters each formula is expanded in.
In order to qualify as an expansion parameter we require that the probability recovers the exact (to all orders) expression as that parameter goes to zero.
That is, $x$ is an expansion parameter if and only if 
\begin{equation}
\lim_{x\to0}P_{\rm approx}(x)=P_{\rm exact}(x=0)\,. 
\label{eq:def-exp-param}
\end{equation}
We note that as many expressions drop higher order terms of more than one parameter at a time, it is quite common for expressions to not be true expansions in the sense of Eq.~\ref{eq:def-exp-param} in that all of the parameters that were treated as small numbers simultaneously.

We find that DMP (at any order) as well as AKT are the only expressions that are an expansion in $s_{13}$ and the matter potential.
Also, while several expressions are expansions in $\eps$ or $\veps$ (including DMP and AKT), some are not an expansion in $\eps$ or $\veps$ either despite treating $\veps$ as a smallness parameter, such as the Madrid-like expressions.
In addition to the parameters listed in the table we note that none of the expression are exact as $L\to0$, the so-called vacuum mimicking regime \cite{Yasuda:2001va}.

\newcommand{\expansion}{{\color{green}\checkmark}}
\newcommand{\none}{{$\color{red}\times$}}
\begin{table}
\centering
\caption{The expansion terms of each expression.
Terms that are expansion parameters in the sense of Eq.~\ref{eq:def-exp-param} are denoted with a green check ({\color{green}\checkmark}), while terms that are not are denoted with a red cross ({$\color{red}\times$}).
Note that Madrid refers also to the expression AJLOS(31) and FL which are all generally quite similar.
AM refers to both AM$^2$ and AM$^{5/2}$, and DMP refers to both DMP$^0$ and DMP$^1$.}
\label{tab:expansion}
\begin{tabular}{|l|c|c|c|}
\hline
             & $\eps$ ($\veps$) & $s_{13}$   & $a/\Delta m^2_{31}$ \\\hline
Madrid(like) & \none            & \none      & \none \\
AKT          & \expansion       & \expansion & \expansion \\
MP           & \expansion       & \none      & \none \\
DMP          & \expansion       & \expansion & \expansion \\
AKS          & \none            & \none      & \none \\
MF           & \expansion       & \none      & \none \\
AJLOS(48)    & \expansion       & \none      & \none \\
AM           & \none            & \none      & \none \\
\hline
\end{tabular}
\end{table}

\subsection{Precision and speed}
In this section we compare the different expressions. As a benchmark point we use the standard oscillation parameters in Tab.~\ref{tab:oscparam}. We take $\delta$ to be slightly off-maximal to avoid any unintentional cancellations and to require both $\sin\delta$ and $\cos\delta$ terms to be correct. We choose as benchmark baseline $L = 1300$ km and density $\rho = 3$ g/cm$^3$, the configuration for the DUNE experiment~\cite{Abi:2018dnh,Abi:2018alz,Abi:2018rgm} although our results are applicable to any current or future long-baseline experiment, including those focusing on the second oscillation maximum. 

\begin{table}
\centering
\caption{Neutrino oscillation parameters used. All of the values are within the 1$\sigma$ ranges of the best fit values obtained in the global fit in Ref.~\cite{deSalas:2017kay}.}
\label{tab:oscparam}        
\begin{tabular}{|c|c|}
\hline
Parameter & Value
\\
\hline
$\Delta m^2_{21}$       & $7.5\times 10^{-5}$ eV$^2$\\  
$\Delta m^2_{ee}$       & $2.50\times 10^{-3}$ eV$^2$\\
$\sin^2\theta_{12}$     & 0.32\\ 
$\sin^2\theta_{23}$     & 0.55\\
$\sin^2\theta_{13}$     & 0.022\\
$\delta$                & -0.40$\pi$\\
\hline
\end{tabular}
\end{table}

\begin{figure}
\centering
\includegraphics[width=0.85\textwidth]{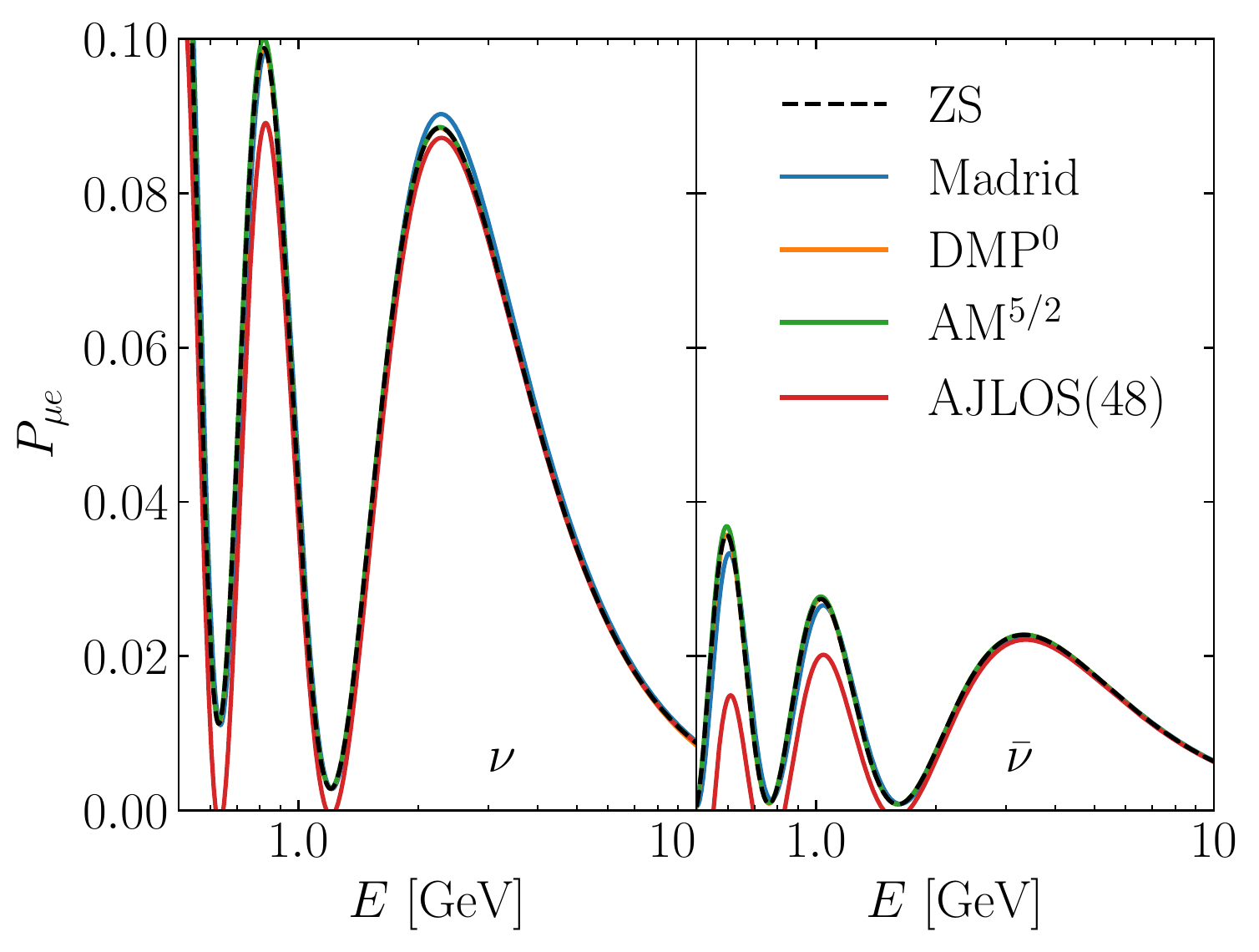}
\caption{Various expressions for the oscillation probabilities for neutrinos (left) and antineutrinos (right) at $L=1300$ km.
The exact expression (ZS) is shown as a black dashed curve. Note that the AM$^{5/2}$ expression (the AM expression to the $5/2$ order) nearly matches the exact expression, although some deviation is visible at the second and third maxima.
The DMP$^0$ (the DMP expression through zeroth order) expression matches the exact expression pixel for pixel in the figure here.
DMP$^1$ is not shown as it is even more precise than DMP$^0$.}
\label{fig:neut-antineut}
\end{figure}

\begin{figure}
\centering
\includegraphics[width=0.85\textwidth]{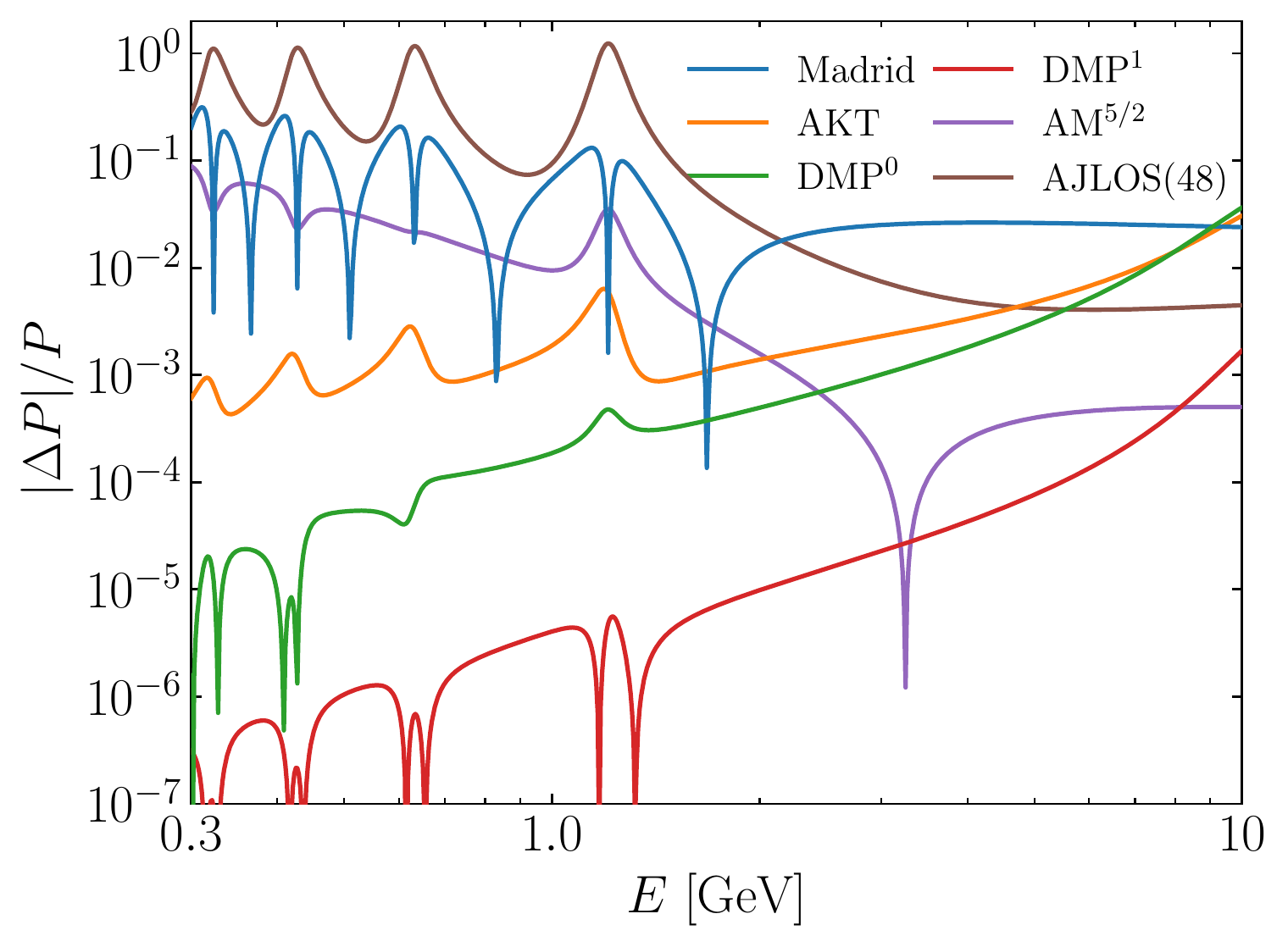}
\caption{The relative precision in the neutrino channel of several of the expressions compared to the exact expression.
The various sharp dips down are where the approximate and exact solutions cross.
Note that towards lower energies all the expressions tend to do poorly except DMP.
At high energies $|\Delta P|/P=|P_{\text{test}}-P|/P$ with $P=P_{\text{ZS}}$ becomes somewhat misleading as $P\to0$, but the precision of each formula levels out (for DMP$^0$ and AKT (DMP$^1$) it levels out at $\Delta P/P=0.057$ ($0.007$) past the atmospheric resonance).}
\label{fig:precision}
\end{figure}

\begin{figure}
\centering
\includegraphics[width=0.85\textwidth]{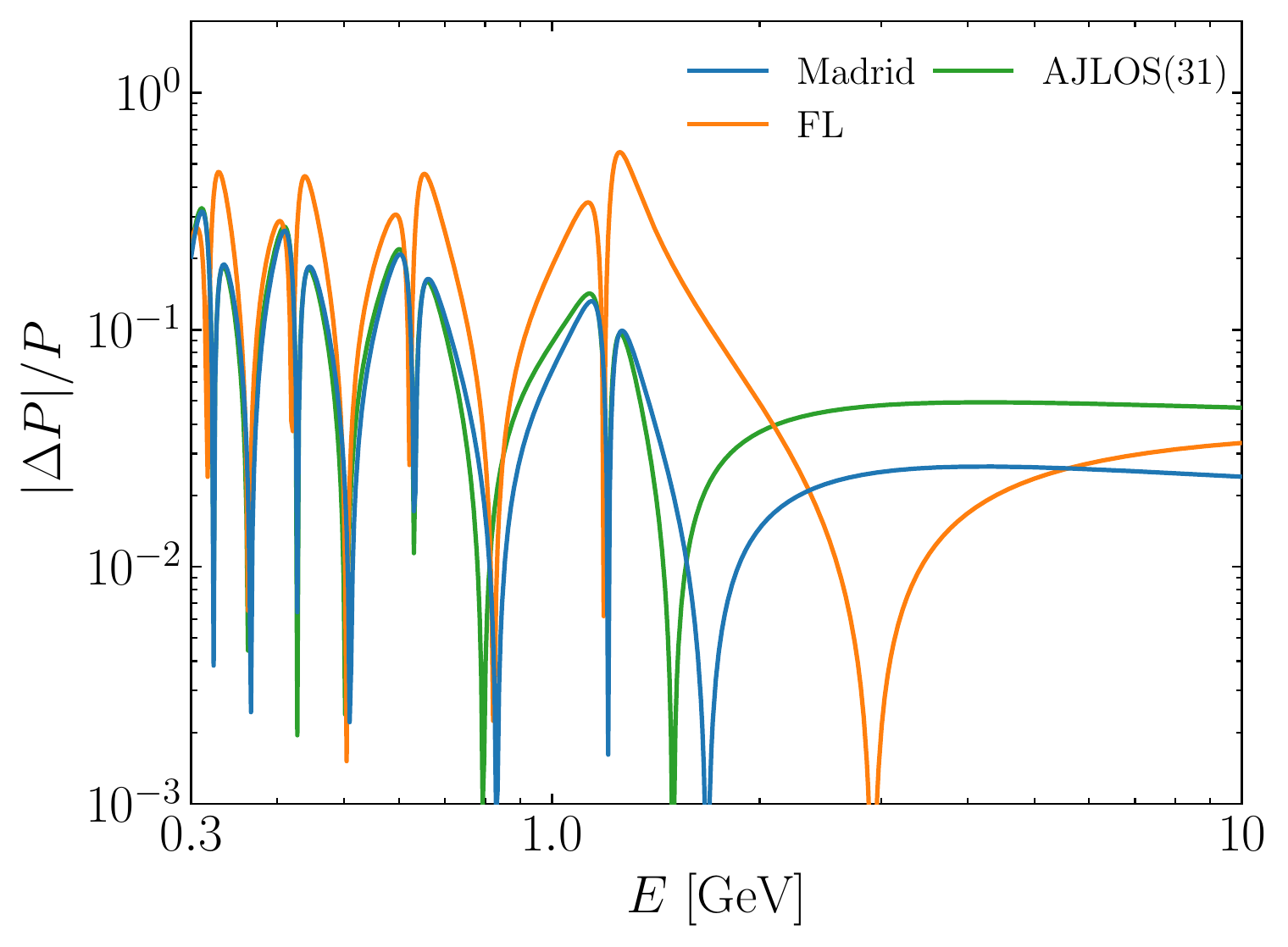}
\caption{The precision of the Madrid-like expressions.}
\label{fig:precision madrid}
\end{figure}

\begin{figure}
\centering
\includegraphics[width=0.85\textwidth]{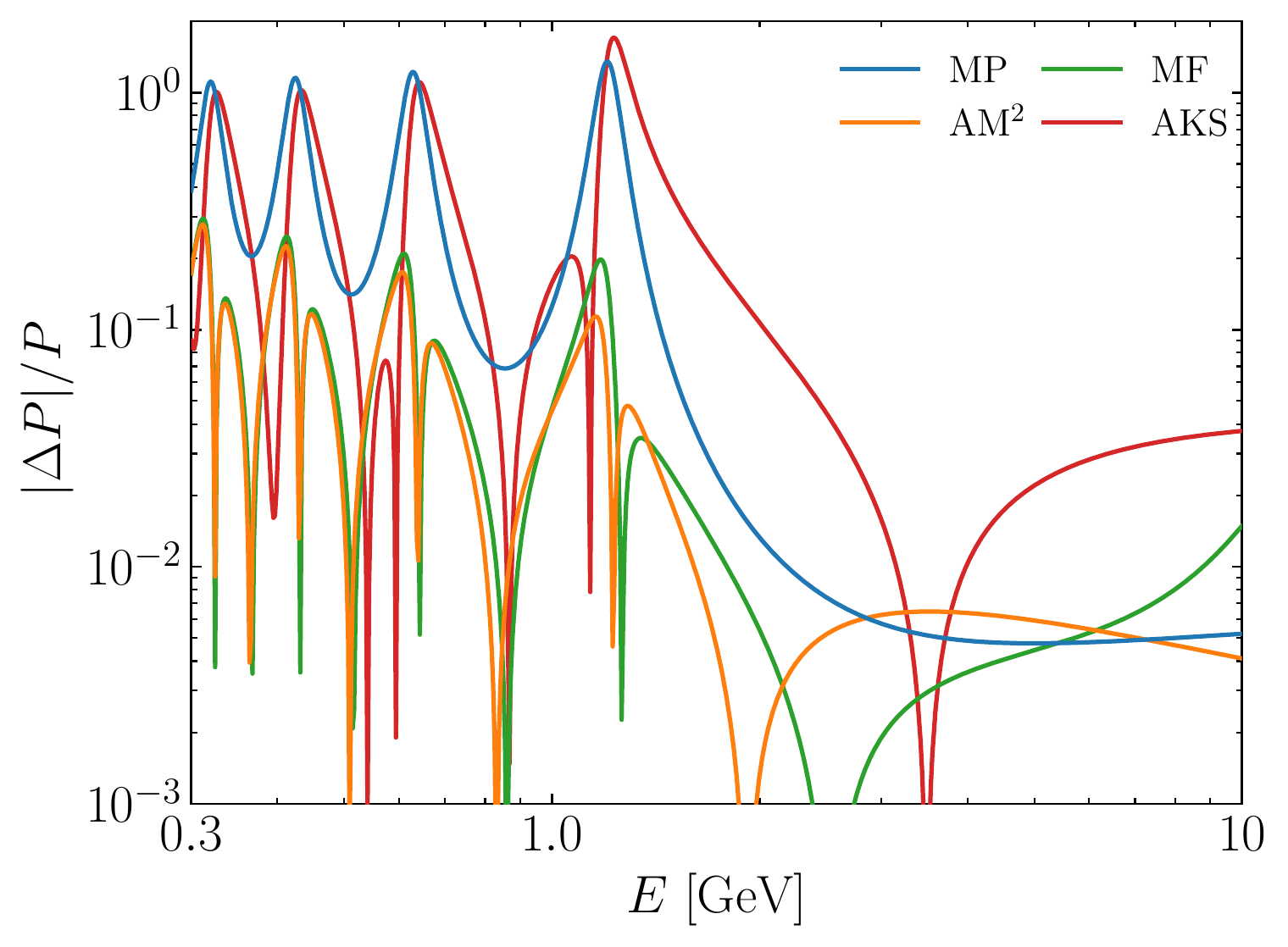}
\caption{The precision of the remaining expressions.}
\label{fig:precision others}
\end{figure}

The probabilities for selected expressions are shown in Fig.~\ref{fig:neut-antineut} for both neutrinos (left) and antineutrinos (right).
They should be compared to the exact curves given by ZS.
To compare the precision of the formulas in a more quantitative way we show $\frac{|P_{\text{test}}-P|}{P}=\frac{|\Delta P|}{P}$, where $P=P_{\text{ZS}}$ is the exact formula. The result is shown in Fig.~\ref{fig:precision}. As one can see for low energies DMP gives the best results, while for $E>2$ GeV the most precise result is AM$^{5/2}$.
The precision of DMP$^0$ and AKT is the worst at the atmospheric resonance ($\sim11$ GeV) before leveling off, although the probability is approaching zero thus this region is less relevant experimentally.
We also show the precision of the several Madrid-like expressions in Fig.~\ref{fig:precision madrid}. The precision of the remaining expressions can be found in Fig.~\ref{fig:precision others}.
Note that the sharp dips are not representative of improved precision, rather they represent a crossing between the exact and approximate expressions.
In addition, the peaks in the errors are when the oscillation probability, and thus the denominator, goes to zero.

\begin{figure}
\centering
\includegraphics[width=0.85\textwidth]{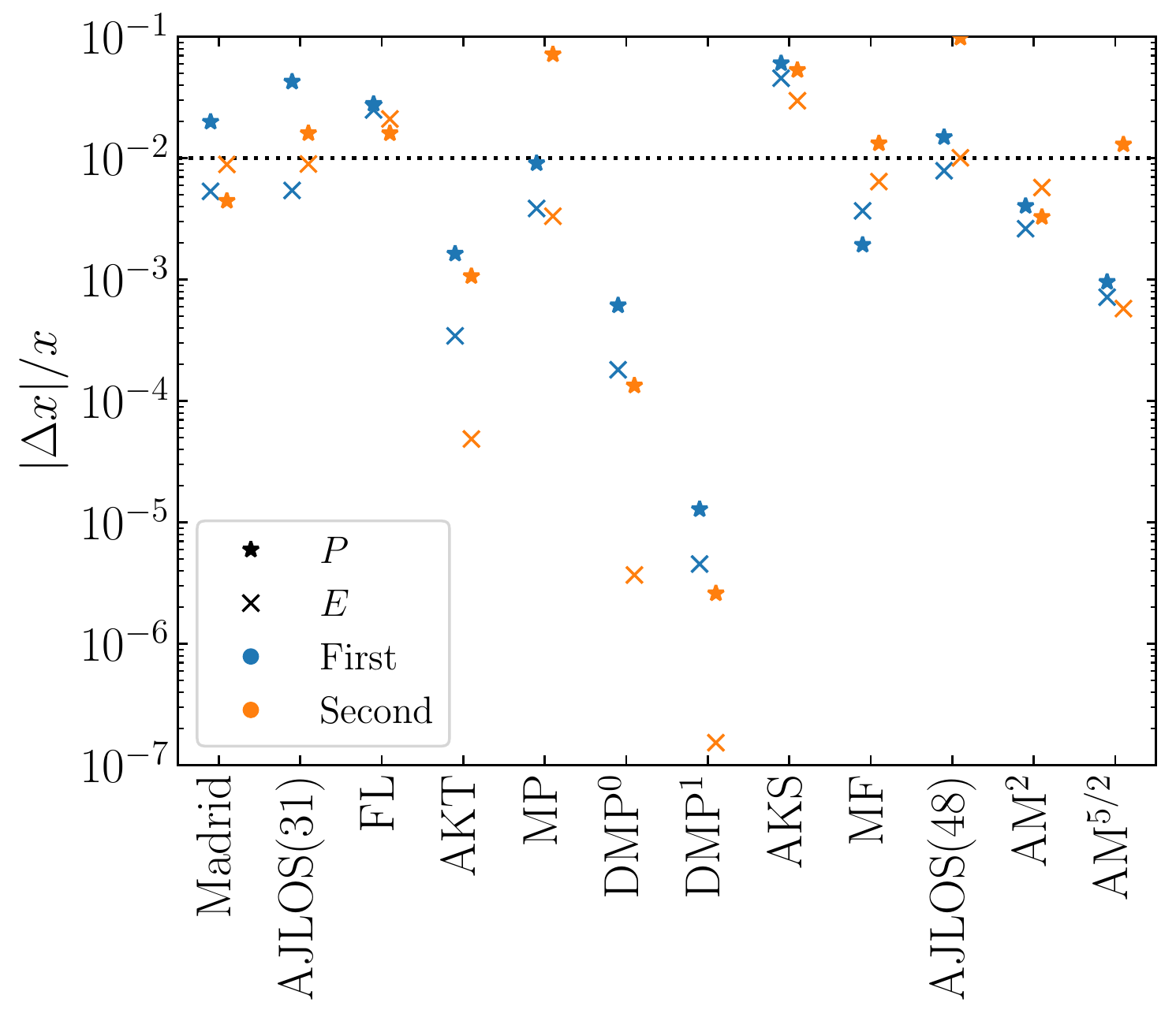}
\caption{The relative error at the first (blue, left) and second (orange, right) oscillation maxima in the probability (star) and energy (cross) for each formula.
While evaluating these on the peaks could lead to a chance conspiracy, we have verified that the results are fairly robust under changes in $\delta$ and in energy away from the maxima.}
\label{fig:peak precision}
\end{figure}

In order to more clearly compare the precision of each expression, we show in Fig.~\ref{fig:peak precision} the precision with which each expression reconstructs the first and second oscillation maxima.
We focus on the oscillation maxima because the heights (probability) of the maxima are an important test for CP violation \cite{Marciano:2001tz} and the locations (energy) of the maxima are an important test for the atmospheric mass splitting \cite{Minakata:2001qm}.
The horizontal line at 1\% is to guide the eye.
Since DUNE and other next generation long-baseline experiments are aiming to reach near the percent level in precision, we cannot introduce theoretical errors larger than 1\%.
We have also verified that these results are generally robust under changes to the oscillation parameters, although for certain specific values of say the CP-violating phase some of the fairly precise expressions may appear to perform much better if there is a crossing between the approximate and exact expressions at the oscillation maximum.

\begin{figure}
\centering
\includegraphics[width=0.85\textwidth]{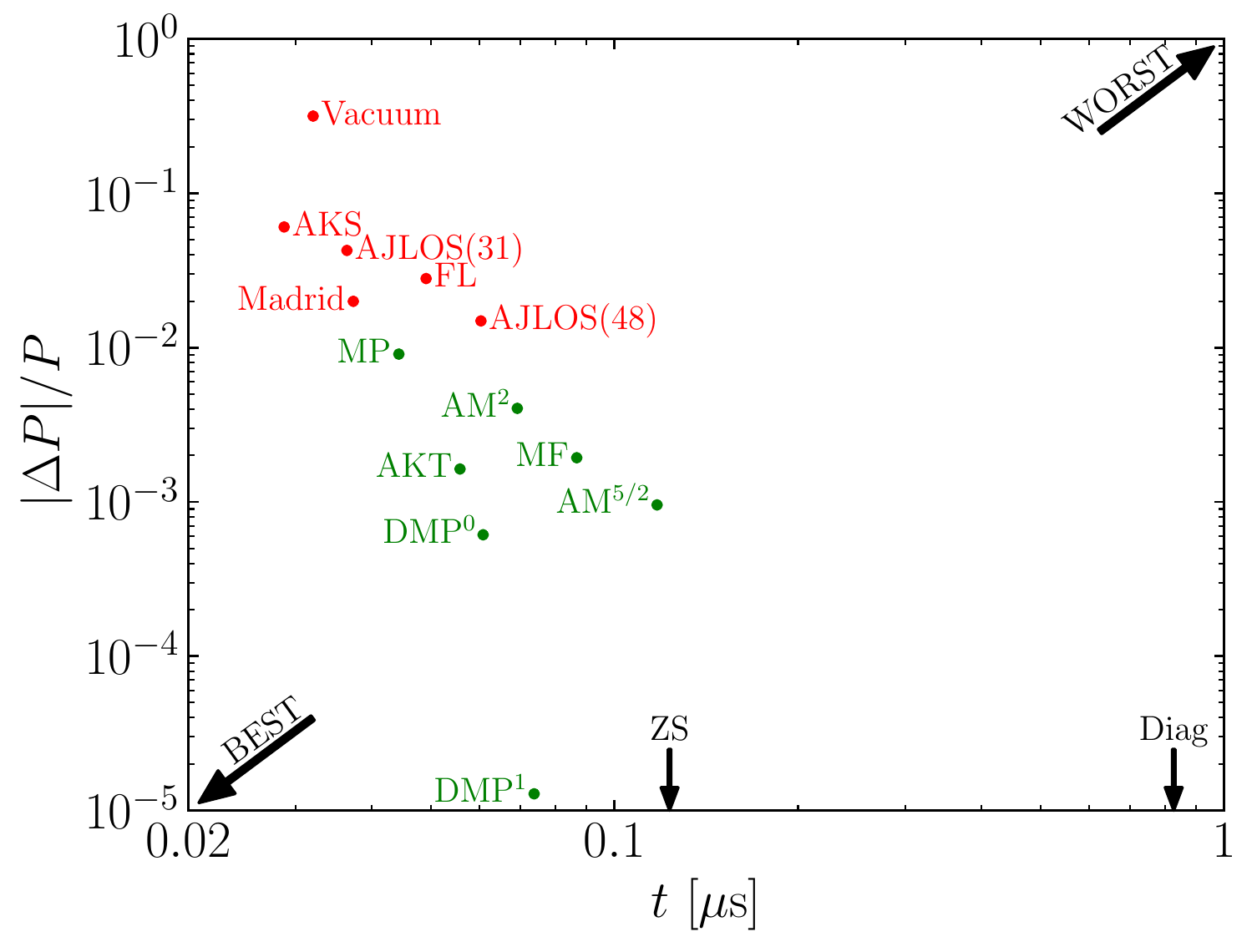}
\caption{The precision of each expression compared with the computational speed.
The vacuum dot refers to calculating the exact vacuum expression without any matter effects.
The precision is defined as a the relative error in probability at the first maximum and the speed is the time it takes to compute one probability.}
\label{fig:speed}
\end{figure}

We also measured the computational time on a single core i7 processor using c++ with the gnu v7.3.0 compiler\footnote{We have performed the same test using Fortran with the gfortran compiler, obtaining qualitatively the same results.}.
Our result can be found in Fig.~\ref{fig:speed} along with the precision at the first oscillation maximum.
The green (red) dots are those for which the probability at the first maximum is better (worse) than 1\%.
We also measure the computational time for the exact vacuum expression from Eq.~\ref{eq:vacuum probability} for comparison.
All of the approximate expressions and ZS are faster than the diagonalization by about an order of magnitude or more, depending on the expansion of interest.
The DMP expressions have the best precision among the approximate expressions by a considerable amount.

When computational speed is the primary, if 1\% precision is required then MP is the best expression, although it is right at $\sim1\%$ and has much larger errors at $\sim1$ GeV.
After MP, the next simplest expression that is also better than 1\% is DMP$^0$ which is precise at the per mille level or better.
The software used to perform these tests \verb1Nu-Pert-Compare1 is publicly available at \href{https://github.com/PeterDenton/Nu-Pert-Compare}{github.com/PeterDenton/Nu-Pert-Compare} \cite{nu-pert-compare}.

\section{Conclusions}
\label{sec:conclusions}
With the advent of next generation long-baseline neutrino experiments, neutrino oscillation physics will truly enter the precision era.
Over the years, various approximate expressions describing oscillation probabilities have been written down.
In this paper we have normalized the notation as much as possible and categorized similar expressions.
We then directly compared them with the exact solution with an eye for precision and speed where the latter provides a rough proxy for simplicity.

We have found that the DMP expressions from Ref.~\cite{Denton:2016wmg} (Denton, Minakata, Parke) are the most precise expressions available in most cases, even at zeroth order, with the option for considerably improved precision at higher orders.
In terms of simplicity they are comparable to any of the other expressions available as shown in Fig.~\ref{fig:speed}.
It is interesting to see within the expression from Ref.~\cite{Asano:2011nj} (Asano and Minakata) that while the addition of the $5/2$ term adds considerably to the precision making it about as precise as any expression, leads to a considerable loss in simplicity and speed as can be seen from Eq.~\ref{Pemu-5/2} and Fig.~\ref{fig:speed}.
In the same vein we can see in Fig.~\ref{fig:speed} that DMP$^1$ adds much more precision (close to two orders of magnitude) than the already quite precise DMP$^0$ with only a modest cost in complexity due to the compact form of the first order correction.

Finally, we note that most of the most precise expressions, AKT and the DMP expressions, all naturally use $\Delta m^2_{ee}$ which is the $\nu_e$ average of $\Delta m^2_{31}$ and $\Delta m^2_{32}$.
This was not ``forced'' or put in by hand, rather it naturally appears.
This suggests that whenever one atmospheric mass splitting is required, the correct expression to use is $\Delta m^2_{ee}$ since there is no reason to prefer one of $\Delta m^2_{31}$ or $\Delta m^2_{32}$ over the other.

\section*{Acknowledgements}
We thank Chris Weaver and Carlos Arg\"uelles for helpful discussions.
GB acknowledges support from  the MEC and FEDER (EC) Grant SEV-2014-0398, FIS2015-72245-EXP, and FPA2014-54459 and the Generalitat Valenciana under grant PRO\-ME\-TEOII/2017/033, also partial support along with SJP from the European Union FP7 ITN INVISIBLES MSCA PITN-GA-2011-289442 and Invisibles\-Plus (RISE) H2020-MSCA-RISE-2015-690575.
PBD acknowledges the United States Department of Energy under Grant Contract desc0012704.
This manuscript has been authored (SJP) by Fermi Research Alliance, LLC under Contract No.~DE-AC02-07CH11359 with the U.S.~Department of Energy, Office of Science, Office of High Energy Physics.
SJP thanks IFT of Madrid for wonderful hospitality during part of this work.
This project has received funding/support from the European Union's Horizon 2020 research and innovation programme under the Marie Sklodowska-Curie grant agreement Nos 690575 and 674896.
CAT  is supported by the Spanish grants FPA2017-90566-REDC (Red Consolider MultiDark), FPA2017-85216-P and SEV-2014-0398 (MINECO/AEI/FEDER, UE), as well as PROMETEO/2018/165 (Generalitat Valenciana) and the FPI fellowship BES-2015-073593. CAT acknowledges the hospitality of the Fermilab Theoretical Physics Department, where part of this work was done. 

\section*{References}
\bibliography{bibliography}

\end{document}